\documentclass[a4paper,reqno,12pt,draft]{article}
\usepackage{amssymb,euscript}

\newcommand{\ie}{{\it i.e.}}
\newcommand{\be}{\begin{equation}}
\newcommand{\ee}{\end{equation}}
\newcommand{\bea}{\begin{eqnarray}}
\newcommand{\eea}{\end{eqnarray}}
\newcommand{\nn}{\nonumber\\}
\newcommand{\p}{\partial}

\newcommand{\Tr}{{\rm Tr}}

\begin{document}

\begin{flushright}
\end{flushright}
\begin{flushright}
\end{flushright}
\begin{center}

{\LARGE {\sc Modeling Multiple M2's}}\\
\bigskip
{\sc Jonathan Bagger\footnote{bagger@jhu.edu}}\\
{ Department of Physics and Astronomy\\
Johns Hopkins University\\
3400 North Charles Street\\
Baltimore, MD 21218, USA
}
\\
\bigskip
and
\bigskip
\\
{\sc Neil Lambert\footnote{neil.lambert@kcl.ac.uk}} \\
{Department of Mathematics\\
King's College London\\
The Strand\\
London WC2R 2LS, UK\\}

\bigskip
{\it Dedicated to the Memory of Andrew Chamblin}
\bigskip

\end{center}

\bigskip
\begin{center}
{\bf {\sc Abstract}}
\end{center}

We investigate the worldvolume theory that describes $N$ coincident
M2-branes ending on an M5 brane. We argue that the fields that
describe the transverse spacetime coordinates take values in a
non-associative algebra.  We postulate a set of supersymmetry
transformations and find that they close into a novel gauge
symmetry.  We propose a three-dimensional $N=2$ supersymmetric
action to describe the truncation of the full theory to the scalar
and spinor fields, and show how a Basu-Harvey fuzzy funnel arises as
the BPS solution to this theory.

\newpage

\section{\sl Introduction}

In \cite{Basu:2004ed} Basu and Harvey  considered the
BPS state of $N$ coincident
M2-branes ending on an M5-brane,
\be
\matrix{M5:&0&1&&3&4&5&6\cr M2:&0&1&2 &&&&\cr}
\label{config}
\ee
On the M5-brane worldvolume, this configuration appears as a
self-dual string soliton \cite{Howe:1997ue}.  Basu and Harvey,
however, examined this configuration from the M2-brane point of
view.  They exploited an analogy with the type IIB string
configuration built from $N$ coincident D1-branes ending on a
D3-brane.  In that case, the end point of the D1-branes appears as a
BPS monopole on the D3-brane worldvolume. On the D1-brane
worldvolume, the configuration gives rise to a `fuzzy-funnel'
soliton \cite{Constable:2001ag}, a fuzzy 2-sphere whose radius grows
without bound as the D3-brane is reached.  These two descriptions of
the same physical state provide a stringy realization of the Nahm
construction
\cite{Diaconescu:1996rk,Kapustin:1998pb,Tsimpis:1998zh}. This leads
one to hope that the M2-brane theory might provide a generalized
Nahm construction for self-dual string solitons.

Basu and Harvey proposed that the M2-brane worldvolume admits a fuzzy-funnel solution that satisfies a generalized Nahm equation,
\be {dX^a\over d(x^2)} + \frac{K}{4!}\epsilon^{abcd}
[G,X^b,X^c,X^d] =0, \label{BHeq} \ee
where $K = M^3/8\pi\sqrt{2N}$ \cite{Basu:2004ed,Berman:2006eu} is a
constant, $a,b,c=3,4,5,6,$
\be
[A,B,C,D] = ABCD - BACD - ACBD+ACDB + ...
\label{nothing}
\ee
and $G$ is a fixed matrix such that  $G^2=1$.
The solution describes a fuzzy 3-sphere \cite{Guralnik:2000pb} whose radius grows without bound as one approaches the M5-brane.  The setup has since be studied from a variety of viewpoints \cite{Berman:2005re,Nogradi:2005yk,Berman:2006eu}.

Unfortunately, it is not known how to derive the Basu-Harvey equation from first principles.  In \cite{Basu:2004ed} a bosonic theory was constructed, essentially by reversing the Bogmoln'yi procedure of writing the action as a perfect square plus boundary terms.  One would like to understand the origin of this theory considering only the geometric and supersymmetric features of M2-branes.  This has not yet been done.

One difficulty stems from the fact that M-theory is the strong
coupling limit of type IIA string theory and hence M2-branes are the
strong coupling limit of D2-branes.  This implies that the
worldvolume theory for $N$ M2-branes is the infra-red fixed point of
a maximally supersymmetric three-dimensional $U(N)$ super-Yang-Mills
theory.

There is no known Lagrangian description of this system.  The only interacting Lagrangian in three dimensions with sixteen supersymmetries is maximally supersymmetric Yang-Mills, which contains one vector plus seven scalars with an $SO(7)$ symmetry.  Simple counting suggests that the M2-brane theory should contain eight scalar fields and an $SO(8)$ symmetry.  In the Abelian case, corresponding to a single M2-brane, such a theory can be obtained directly from the D2-brane worldvolume theory by dualizing the vector field into a scalar.  In the non-Abelian case, however, there is no straightforward way to do this.

There are other peculiar features of the multi M2-brane system that are difficult to reconcile with quantum field theory based on a Lagrangian.  For example, the near horizon limit of $N$ M2-branes is dual to a three-dimensional CFT with $N^{\frac{3}{2}}$ degrees of freedom.  Also, the lack of a free parameter in M-theory implies that there is no free parameter in the M2-brane worldvolume theory.  This suggests that there is no weakly coupled limit that might be described by perturbative quantization of a classical Lagrangian.

Despite these difficulties, it is still of interest to try to
construct a classical theory that can capture at least some of the
features of multiple M2-branes.  One might be able to identify the
field content and supersymmetry transformations, and give a
geometrical interpretation to the fields.  A recent attempt was made
in \cite{Schwarz:2004yj} (see also \cite{Lee:2006hw}), where the
scalar fields were taken to be $U(N)$-valued, transforming under a
standard gauge symmetry. The gauge field kinetic term was taken to
be of Chern-Simons type, so the vector field did not introduce any
propagating degrees of freedom.  Under these assumptions, no theory
was found with sixteen supersymmetries.

In this paper we present an alternative approach to the multi
M2-brane system.   We propose a classical Lagrangian for the
scalar-spinor sector of the theory.  Our Lagrangian is
supersymmetric and scale invariant with manifest $SU(4)\times U(1)
\in SO(8)$ symmetry. We recover the Basu-Harvey equation at the cost
of introducing a non-associative algebra for the eight coordinates,
or more correctly, an algebra for which the Jacobi identity is not
satisfied.  (A slightly different role for non-associative algebras
in the Basu-Harvey equation was discussed in \cite{Berman:2006eu}.)
We will see, as a consequence of our assumptions, that the full
M2-brane theory must contain a rather curious gauge symmetry, one
that we will not study in detail in this paper.

The rest of this paper is organized as follows. In the next section we describe
the supersymmetry transformations of multiple D2- and M2-branes, and argue that the M2-brane coordinates are naturally elements of a non-associative algebra.  In section three we compute the closure of the supersymmetry transformations and find evidence for a novel gauge symmetry associated with the M2-branes.  In section four we propose a Lagrangian with four supersymmetries that might model some features of the complete multiple M2-brane theory.  In section
five we find a BPS solution and show how the Basu-Harvey equation arises in a simple example.
In the final section we state our conclusions and present directions for future work.

\section{\sl Supersymmetry Transformations}

We start by considering the supersymmetry transformations of $N$ coincident D2-branes, written so the spacetime symmetries are manifest:
\bea
\delta X^i &=& i\bar\epsilon\Gamma^i\Psi\nn \delta A_\mu &=&
i\bar\epsilon \Gamma_\mu\Gamma^{10}\Psi\nn \delta \Psi &=& \p_\mu
X^i\Gamma^\mu \Gamma^i\epsilon +
\frac{1}{2}F_{\mu\nu}\Gamma^{\mu\nu}\Gamma^{10}\epsilon
+\frac{i}{2}[X^i,X^j]\Gamma^{ij}\Gamma^{10}\epsilon.
\eea
Here $\mu,\nu=0,1,2$ label the worldvolume coordinates, and $i,j=3,...,9$ label the transverse dimensions of the D2-branes.  There is an $SO(1,2)$ symmetry of the worldvolume, as well as a manifest $SO(7)$ symmetry of the transverse $\mathbb{R}^7$ that acts on the scalars and on the $\Gamma$ matrices.  Notice the explicit appearance of  $\Gamma^{10}$.  This matrix ensures that the unbroken supersymmetries satisfy
\be
\Gamma^{012}\epsilon = \epsilon,
\ee
while the broken supersymmetries satisfy
\be
\Gamma^{012}\epsilon = -\epsilon.
\ee
All the Fermions are Goldstinos, and obey the corresponding parity
condition,
\be
\Gamma^{012}\Psi = -\Psi.
\ee

We now attempt to generalize these transformations to the case of multiple M2-branes.  The presence of the explicit $\Gamma^{10}$ forbids a straightforward lift to eleven dimensions.  Therefore we simply assume that there is {\it some} extension of the D2-brane transformations such that, if all the vector fields are set to zero, the D2-brane transformations lift in such a way that the $SO(7)$ symmetry is trivially extended to $SO(8)$.  Our transformations capture the fact that the M2-brane theory almost certainly contains eight scalar fields, corresponding to the eight transverse dimensions.  Since we have ignored all gauge fields, we cannot expect the corresponding Lagrangian to be invariant under the supersymmetry transformations.  Nor can we expect the transformations to close, and indeed as we shall see, they do not.

Thus in what follows we put aside all vector fields and study the
scalar-spinor supersymmetry transformations of the multi M2-brane
theory.  We propose lowest-order supersymmetry transformations of the following form:
\bea
\delta X^I &=& i\bar\epsilon\Gamma^I\Psi\nn \delta \Psi &=& \p_\mu
X^I\Gamma^\mu \Gamma^I\epsilon
+{i}\kappa[X^I,X^J,X^K]\Gamma^{IJK}\epsilon, \label{susy}
\eea
where $I,J,K = 3,4,5,...,10$.  In these expressions, $\kappa$ is a dimensionless constant and $[X^I,X^J,X^K]$ is antisymmetric and linear in each of the fields.  These transformations imply that $(X^I,\Psi)$ have dimension $(\frac{1}{2},1)$, as required for conformal invariance.  We note that there could be other cubic terms that are not totally anti-symmetric in $I,J,K$ and that vanish in the D2-brane limit, or that correspond to higher-order terms in the Dirac-Born-Infeld effective theory of the D2-branes.  We do not consider these possibilities here.  Instead, we just stipulate the presence of a $\Gamma^{IJK}$ term, and we focus on it alone.

There is a second argument for such a $\Gamma^{IJK}$ term in the supersymmetry transformations.  The preserved supersymmetries of $N$ M2-branes in the presence of an M5-brane satisfy $\Gamma^{2}\epsilon= \Gamma^{3456}\epsilon$, or equivalently
\be
\Gamma^{abc}\epsilon = \epsilon^{abcd}\Gamma^2\Gamma^{d}\epsilon,
\label{eproj}
\ee
where $a,b,c,d=3,4,5,6$.  From this one obtains the BPS equation
\be \frac{dX^a}{ d(x^2)} = i\kappa\epsilon^{abcd}[X^b,X^c,X^d] .
\label{BPS} \ee
The solutions to this equation behave as $X^a \sim 1/\sqrt{x^2}$ as $X^a \to\infty$.  Turning this around, we see that $x^2 \sim 1/R^2$ at small $R$, where $R^2= (X^3)^2+(X^4)^2+(X^5)^2+(X^6)^2$.  This is the correct divergence to reproduce the profile of the self-dual string soliton on the M5-brane \cite{Howe:1997ue}.  The cubic term and the appearance of the $\Gamma^{IJK}$ are crucial to obtaining a Bogomoln'yi equation with the correct features.

The scalar fields are valued in an algebra ${\cal A}$.  Translational invariance requires the algebra to have a centre, an element $I$ that commutes with everything,
\be
X^I \to X^I + x^I I.
\ee
One's first impulse is to take the $X^I$ to be valued in the Lie algebra $\underline{u}(N)$, as in the D2-brane theory.  The $[X^I,X^J,X^K]$ would then be given by a double commutator,
\be
[X^I,X^J,X^K] = \frac{1}{3!}[X^{I},X^J],X^{K}] \pm {\rm cyclic},
\ee
which vanishes because of the Jacobi identity.  Therefore in what follows we take the $X^I$ to be valued in a non-associative algebra ${\cal A}$, with a product $\cdot :{\cal A} \times {\cal A}\to {\cal A}$.  We require the algebra to have a one-dimensional centre generated by $I$, and define the associator
\be
<X^I,X^J,X^K> = (X^I\cdot X^J)\cdot X^K-X^I\cdot (X^J\cdot X^K).
\ee
We then define
\be
[X^I,X^J,X^K]= \frac{1}{2\cdot 3!}<X^{[I},X^J,X^{K]}>,
\ee
which is linear and fully antisymmetric, as required.  With
this construction, we have defined the supersymmetry
transformations (\ref{susy}) for the scalar-spinor sector of
the M2-brane theory.

\section{\sl Closure}

In the previous section, we argued that the M2-brane supersymmetry transformations cannot be expected to close because we had set all gauge fields to zero.  Indeed, commuting the transformations (\ref{susy}), we find
\bea [\delta_1,\delta_2]X^I &=& 2i\bar\epsilon_1\Gamma^\mu\epsilon_2
\p_\mu X^I +6\kappa\bar\epsilon_1\Gamma^{JK}\epsilon_2[X^I,X^J,X^K]
\nn {}[\delta_1,\delta_2]\Psi &=&
2i\bar\epsilon_1\Gamma^\mu\epsilon_2 \p_\mu \Psi
-i\bar\epsilon_1\Gamma^\mu\epsilon_2 \Gamma^\mu
(\Gamma^\lambda\p_\lambda\Psi+\frac{9i\kappa}{4}\Gamma^{IJ}[X^I,X^J,\Psi]
) \nn &&-\frac{i}{4}\bar\epsilon_1\Gamma^{KL}\epsilon_2 \Gamma^{KL}
(\Gamma^\mu\p_\mu\Psi+3i\kappa\Gamma^{IJ}[X^I,X^J,\Psi] )\nn
&&-\frac{9\kappa}{8}\bar\epsilon_1\Gamma^{IJ}\epsilon_2[X^I,X^J,\Psi]
-6\kappa\bar\epsilon_1\Gamma^{IK}\epsilon_2\Gamma^{JK}[X^I,X^J,\Psi]
\nn &&+\frac{3\kappa}{8\cdot
4!}\bar\epsilon_1\Gamma^\lambda\Gamma^{KLMN}\epsilon_2
\Gamma^\lambda\Gamma^I\Gamma^{KLMN}\Gamma^J[X^I,X^J,\Psi]. \eea
The first term on the right hand side of each variation is a
translation, generated by the vector
$\bar\epsilon_1\Gamma^\mu\epsilon_2$.  The second term on the right
hand side of $[\delta_1,\delta_2]\Psi$ is generated by the same
vector, but it is not a translation. Therefore it seems plausible to
remove it by assuming that the spinor equation of motion is
\be
\Gamma^\mu\p_\mu\Psi+\frac{9i\kappa}{4}\Gamma^{IJ}[X^I,X^J,\Psi]=0.
\ee
With this assumption, the supersymmetries close on world-sheet translations and the following set of bosonic transformations
\bea
\delta X^I &=&6i\kappa v_{JK}[X^I,X^J,X^K] \nn \delta \Psi &=&
\frac{3i\kappa}{16}v_{KL} \Gamma^{KL}\Gamma^{IJ}[X^I,X^J,\Psi]\nn
&&-\frac{9i\kappa}{8}v_{IJ}[X^I,X^J,\Psi] -6i\kappa v_{IK}\Gamma^{J
K}[X^I,X^J,\Psi] \nn &&+\frac{3i\kappa}{8\cdot 4!}v_{\lambda KLMN}
\Gamma^\lambda\Gamma^I\Gamma^{KLMN}\Gamma^J[X^I,X^J,\Psi] ,
\label{noclose}
\eea
where
\be v_{IJ}=-i\bar\epsilon_1\Gamma_{IJ}\epsilon_2\qquad{\rm
and}\qquad v_{\lambda KLMN} =
-i\bar\epsilon_1\Gamma_\lambda\Gamma_{KLMN}\epsilon_2.
\label{noclose2}
\ee

These transformations are very mysterious.  In the analogous
calculation for multiple D2-branes, the supersymmetry algebra closes
on translations and local gauge transformations. Therefore one is
tempted to associate these terms with some type of generalized gauge
transformations on the M2-branes.  However, it is difficult to
understand how such a transformation can arise in the
strong-coupling limit of the D2-brane theory.  The situation might
be clarified by including appropriate gauge fields, but without
additional Fermions, the gauge fields cannot contain any propagating
degrees of freedom.

One can speculate that the non-closure of the multi M2-brane algebra might, in fact, be related to the presence of the M5-brane.  To see why, let us introduce the following closed 5-form
\be
\omega_{mnpqr} = -\frac{i}{5!}\bar\epsilon_1\Gamma_{mnpqr}\epsilon_2
\ee
in eleven dimensions, where $m,n,...=0,...,10$.  Because
$\Gamma_{012}\epsilon_{1,2} =\epsilon_{1,2}$, the non-vanishing
components of $\omega$ are precisely the forms generated by the
closure (\ref{noclose}), (\ref{noclose2}):
\be
\omega_{\mu\nu\lambda IJ} = \epsilon_{\mu\nu\lambda}v_{IJ}\ ,\qquad
\omega_{\lambda KLMN} = v_{\lambda KLMN}.
\ee
Furthermore, $\omega_{mnpqr}$ is the calibrating form
associated with an M5-brane that lies in the $x^m,...,x^r$ plane.
This suggests that the multiple M2-branes might
somehow be acting as a source for the M5-brane.

In what follows, we construct a toy Lagrangian for the M2-brane system that is invariant under four of the sixteen supersymmetries.  We start by adopting a spinor notation that is better suited to three dimensions. We also use a notation in which only an $SU(4)\times U(1)$ symmetry of the transverse $SO(8)$ is manifest.  With these conventions, we define the fields to transform as follows,
\be
\matrix{X^I &\to& Z^A \oplus Z_{\bar A} &\in & {\bf 4}(1) \oplus {\bar {\bf
4}}(-1)\cr
\Psi &\to &\psi^A \oplus \psi_{\bar A} &\in &{\bf 4}(-1)
\oplus {\bar {\bf 4}}(1)\cr
\epsilon&\to&
\varepsilon\oplus\varepsilon^{*}\oplus \varepsilon^{AB} &\in & {\bf
1}(-2) \oplus {\bar {\bf 1}}(2) \oplus {\bf 6}(0). \cr}
\ee
In this notation, the transformations (\ref{susy}) take the form
\bea
\delta Z^A &=& i\bar\varepsilon \psi^A +
i\bar\varepsilon^{AB}\psi_{\bar B}\nn \delta \psi^A &=&
2\gamma^\mu\p_\mu Z^A \varepsilon +2\gamma^\mu
\varepsilon^{AB}\p_\mu Z_{\bar B}  +
i\kappa_1\epsilon^{ABCD}[Z_{\bar B},Z_{\bar C},Z_{\bar
D}]\varepsilon^*\nn && +3i\kappa_2[Z^{A},Z_{\bar B},Z_{\bar
C}]\varepsilon^{BC} +3i\kappa_3[Z^{ A},Z^{ B},Z_{\bar
B}]\varepsilon,
\eea
where the dimensionless parameters $\kappa_1$, $\kappa_2$ and $\kappa_3$ are all non-zero and proportional to $\kappa$.

Let us restrict our attention to the supersymmetries generated by $\varepsilon$, \ie\ we consider the $N=2$ subalgebra defined by $\varepsilon^{AB}=0$,
\bea
\delta Z^A &=& i\bar\varepsilon \psi^A \nn \delta \psi^A &=&
\gamma^\mu\p_\mu Z^A \varepsilon+ i\kappa_1\epsilon^{ABCD}[Z_{\bar
B},Z_{\bar C},Z_{\bar D}]\varepsilon^* +3i\kappa_3[Z^{ A},Z^{
B},Z_{\bar B}]\varepsilon.\nn \label{xyz}
\eea
This corresponds to imposing
\be
\Gamma_{5678}\epsilon =\Gamma_{56910}\epsilon=-\epsilon
\ee
on the full eleven-dimensional spinor $\epsilon$.  Geometrically this projection arises from additional M-branes along the $01234810$ and $0123589$ planes (as well as other possible M5-branes that can be introduced without breaking additional supersymmetries). This situation corresponds to the  $SU(3)$ Kahler calibration case studied in \cite{Berman:2005re}.

Computing the closure of this subalgebra, we find
\bea
[\delta_1,\delta_2]Z^A &=& 2v^\mu\p_\mu Z^A +i\kappa_3 u[Z^{ A},Z^{
B},Z_{\bar B}]\nn\ [\delta_1,\delta_2]\psi^A &=& 2v^\mu\p_\mu \psi^A
+\frac{i\kappa_3}{2}w^\mu\gamma_\mu [Z^A,Z^B,\psi_{\bar B}]\nn &&
+\frac{i\kappa_3}{2}(u+v^\nu\gamma_\nu)([\psi^A,Z^B,Z_{\bar
B}]+[Z^A,\psi^B,Z_{\bar B}])\nn &&
+\frac{1}{2}(u-v^\nu\gamma_\nu)(2\gamma^\mu\p_\mu\psi^A -
3i\kappa_1\epsilon^{ABCD}[Z_{\bar B},Z_{\bar C},\psi_{\bar D}]),
\label{Neq2}\eea
where
\bea
v^\mu
&=&i\bar\varepsilon_2\gamma^\mu\varepsilon_1-i\bar\varepsilon_1\gamma^\mu\varepsilon_2\nn
u &=&i\bar \varepsilon_2\varepsilon_1-i\bar
\varepsilon_1\varepsilon_2\nn w^\mu &=&i\bar
\varepsilon_2^*\gamma^\mu\varepsilon_1-i\bar
\varepsilon_1^*\gamma^\mu\varepsilon_2.
\eea
The last term in (\ref{Neq2}) is the equation of motion; it vanishes on shell.  The terms proportional to $\kappa_3$, however, prohibit closure of the algebra even when restricted to the above supersymmetries.

The transformations generated by $\varepsilon$ are very similar to those found in three-dimensional $N=2$ superspace.  The only exceptions are the $\kappa_3$ terms, which contain both holomorphic and anti-holomorphic fields.  In particular, let us consider the three-dimensional $N=2$ chiral superfield ${\cal Z}^A$, with
\be
\bar D{\cal Z}^A =0,
\ee
which can be expanded as
\be
{\cal Z}^A = Z^A(y) +\bar\theta^*\psi^A(y) + \bar\theta^*\theta
F^A(y)
\ee
where $y^\mu = x^\mu + i\bar\theta\gamma^\mu\theta$.  We take a real basis for the three-dimensional Clifford algebra; the spinors $\varepsilon$ and $\theta$ are complex with $\bar \theta = \theta^{*T}\gamma^0$.  In terms of components, we find
\bea
\delta Z^A &=& i\bar\varepsilon\psi^A\nn \delta F^A &=&
-\bar\varepsilon^*\gamma^\mu\p_\mu \psi^A\nn \delta \psi^A &=&
2\gamma^\mu\p_\mu Z^A\varepsilon + 2iF^A\varepsilon^*.
\label{2dsusy}
\eea
Comparing (\ref{2dsusy}) with (\ref{xyz}), we see that the transformations coincide when $[Z^A,Z^B,Z_{\bar B}]=0$ and
\be
F^A = \kappa_1\epsilon^{ABCD}[Z_{\bar B},Z_{\bar C},Z_{\bar D}].
\ee
Moreover, the algebra (\ref{Neq2}) closes provided $[Z^A,Z^B,Z_{\bar B}]$, $[Z^A,Z^B,\psi_{\bar B}]$ and $[\psi^A,Z^B,Z_{\bar B}]+[Z^A,\psi^B,Z_{\bar B}]$ are all zero.  These conditions all arise from the single superspace constraint
\be
[{\cal Z}^A,{\cal Z}^B,{\cal Z}_{\bar B}]=0.
\label{susyconstraint}
\ee
Therefore the conjectured M2-brane superalgebra can be truncated to a consistent $N=2$ superalgebra with $SU(4)\times U(1)$ R-symmetry when the superspace constraint (\ref{susyconstraint}) is satisfied.

In closing this section, we observe that the constraint
(\ref{susyconstraint}) is reminiscent of the Gauss law constraint in
ordinary gauge theory.  For example, if we consider the purely
scalar-spinor sector of $N=4$ super-Yang-Mills, we can expect to
see at most $N=1$ supersymmetry and an $SU(3)\times U(1)$ part of
the $SO(6)$ $R$-symmetry.  Furthermore, since the scalars and
spinors act as sources for the gauge field, we find constraints
that come from imposing that these sources vanish.  Therefore we
conjecture that the constraint (\ref{susyconstraint}) has a similar
interpretation, ensuring that the scalar and spinors do not provide
a source for the mysterious gauge fields that we have set to zero.

\section{\sl Superspace Lagrangian}

In the previous section, we considered M2-brane dynamics when all gauge fields are set to zero, only the $R$-charge $\pm 2$ supersymmetries are realized, and the $SO(8)$ $R$-symmetry is broken to $SU(4)\times U(1)$.  In this section we construct an $N=2$ superspace Lagrangian that reproduces many of the features discussed above.

As before, we assume there are eight scalar fields, which we write as four complex scalars $Z^A$, $A=1,2,3,4$.
We require that the $Z^A$ take values in a non-associative algebra ${\cal A}$ with centre $I$.
To write the action, we assume that the algebra is equipped with a trace form, a bilinear operation $\Tr:\ {\cal A}\times{\cal A} \to {\bf \mathbb{C}}$ that satisfies
\be \Tr(A,B) = \Tr(B, A),
\label{Trprod}
\ee
with the invariance condition
\be
\Tr(A\cdot B, C)=\Tr(A, B\cdot C).
\label{Trass}
\ee
This latter condition turns out to be important because it allows us
to evaluate derivatives of the Lagrangian with respect to elements
of $\cal A$.

We also need to define how complex conjugation acts on the algebra.  Therefore we suppose that there is an involution $\#:\,{\cal A} \to{\cal A}$ such that $\#^2=1$ and
\be
\Tr(A,A^\#) \ge 0
\ee
for all $A\in {\cal A}$, with equality if and only if $A=0$.  We then define complex conjugation in the algebra to be given by $\#$, \ie\ if $Z^1 = X^3+iX^7$ then $Z_{\bar 1} = (X^3)^\#-i(X^7)^\#$.

With these definitions, the $N=2$ supersymmetric Lagrangian takes the following form,
\be
{\cal L}  =  \frac{1}{2}\int d^4\theta\, \Tr({\cal Z}^A, {\cal
Z}_{\bar A} ) + \int d^2\theta\, W({\cal Z}^A) + \int d^2\theta^*\, \bar
W({\cal Z}_{\bar A}),
\label{susyL}
\ee
where $W$ is a holomorphic function on the algebra.  We observe that, even though the non-associativity complicates matters such as Taylor expansions, supersymmetry still holds because $\bar\varepsilon Q+\bar\varepsilon^* Q^*$ is a differential operator, even in this more general setting.  The field $F^A$ is auxiliary; its equation of motion is
\be
F_{\bar A} = -2\p_{A} W.
\ee
As noted in \cite{Schwarz:2004yj}, the most natural form for the superpotential is
\be
W \sim \epsilon_{ABCD}\Tr({\cal Z}^A{\cal
Z}^B{\cal Z}^C{\cal Z}^D).
\ee
In three dimensions, a quartic superpotential leads to a cubic term in the supersymmetry transformations and an action that is classically scale invariant.

If the ${\cal Z}^A$ were Lie-algebra valued, the superpotential $W$
would vanish.  For our algebra, however, the superpotential does not
vanish.  In particular, we take
\be
W = -\frac{\kappa_1}{8} \epsilon_{ABCD}\Tr({\cal Z}^A, [{\cal
Z}^B,{\cal Z}^C,{\cal Z}^D]).
\ee
Using (\ref{Trprod}) and (\ref{Trass}), we compute $\p W/\p {\cal Z}^A$ and find
\be
\frac{\p W}{\p {\cal Z}^A} =
-\frac{\kappa_1}{2}\epsilon_{ABCD}[{\cal Z}^B,{\cal Z}^C,{\cal
Z}^D].
\ee
Combining this with (\ref{2dsusy}), we obtain the superalgebra (\ref{xyz}), after imposing the constraint (\ref{susyconstraint}).

It was noted in  \cite{Schwarz:2004yj} that the coefficient of the
superpotential is arbitrary from the point of view of $N=2$
supersymmetry, but that it might be fixed once the full
supersymmetry is realized.  This is also reminiscent of
four-dimensional super-Yang-Mills where the coefficient of the
superpotential is fixed by the full $N=4$ supersymmetry.  Certainly
the multiple M2-brane theory is not expected to have any free
parameters.

It is useful to compare our construction to the case of $SU(N)$
gauge theory. Since the scalar fields are spacetime coordinates,
they are represented by Hermitian matrices.  The reality condition
is preserved by the commutator but not by the matrix product.  More
formally, one can say that starting from an associative $N\times N$
matrix algebra, one constructs the Lie algebra of Hermitian matrices
that is closed under the anti-symmetric product: $i[\ ,\ ]:{\cal
A}\times {\cal A}\to {\cal A} $.

In our case, we start with a non-associative algebra. Since the coordinates are physical, it seems natural to impose a generalized Hermitian condition
\be (X^I)^\#=X^I.
\label{Hconstraint}
\ee
In general, this condition is not  preserved by the algebra product.
However, all we really need is that the triple product be Hermitian,
\be
(i[X^I,X^J,X^K])^\#  =  i [X^I,X^J,X^K].
\ee
Thus, rather than require a Lie algebra with an anti-symmetric bi-linear product, we demand a new algebraic structure with an anti-symmetric tri-linear map $i[\ ,\ ,\ ]:{\cal A}\times {\cal A}\times {\cal A}\to {\cal A}$ that preserves the Hermitian condition.  We
refer to such a structure as a three-algebra.

Finally, let us consider the global symmetries of this Lagrangian.
In the familiar case of $U(N)$ gauge theory, where $\#=\dag$, the
Lagrangian is invariant under $Z^A \to g Z^A g^{-1}$, provided that
$g^{-1}=g^\dag$, \ie\ for $g \in U(N)$. For a non-associative
algebra it is not clear that there is an associated group of which
$g$ is an element.  Furthermore the expression $g Z^A g^{-1}$ is
ambiguous. Nevertheless, for the purposes of field theory, it is
often sufficient to consider transformations $g=1+h$, where $h$ is
an infinitesimal element of the algebra. Thus we can look for
symmetries of the form $\delta Z^A = h\cdot Z^A - Z^A\cdot h$, where
$h^\#=-h$ to preserve the condition $(Z^A)^\#=Z^A$.  One readily
sees that this is always a symmetry of the kinetic term but that
{\it a priori} it is not a symmetry of the superpotential.  Thus in
general there is no global symmetry associated with the algebra that
might be considered as a remnant of a standard gauge symmetry.

To summarize, we have constructed a Lagrangian with three-dimensional $N=2$ supersymmetry (\ie\ four supercharges) and an $SU(4) \times U(1)$ R-symmetry.  The supersymmetry transformations coincide with our conjectured M2-brane transformations (\ref{Neq2}) when the constraint (\ref{susyconstraint}) is satisfied.  Ideally we would have liked a system with $N=8$ supersymmetry and $SO(8)$ R-symmetry, but at least $SU(4)\times U(1)$ is a maximal supergroup of $SO(8)$.  It is conceivable that our Lagrangian admits additional supersymmetries that are not manifest in the superspace formalism, but we have been unable find them.

\section{\sl BPS Funnels}

The supersymmetric Lagrangian (\ref{susyL}) gives rise to the following BPS condition for its bosonic solutions,
\be
0=2\gamma^\mu\p_\mu  Z_{\bar A} \varepsilon^* +4i\p_A W\varepsilon.
\ee
Writing $\varepsilon = e^{i\alpha}\gamma^2\varepsilon^*$, where
$e^{i\alpha}$ is a phase, we see that half the
supersymmetries are preserved whenever
\be
{d Z_{\bar A}\over d (x^2)} - i\kappa_1 e^{i\alpha}
\epsilon_{ABCD}[{Z}^B,{Z}^C,{ Z}^D]=0.
\ee
This is the BPS equation (\ref{BPS}).

The energy density of a BPS configuration extended along the $x^1$
direction is
\bea
E  &=& \frac{1}{2}\int dx^2  \Tr\,\p_2(Z^A,\p_2 Z_{\bar A}) +
4\Tr\,(\p_A W,\p_{\bar A} \bar W)\nn &=&\frac{1}{2} \int dx^2
\Tr\,(\p_2Z^A - 2ie^{-i\alpha}\p_{\bar A} \bar W,\p_2Z_{\bar A}+2i
e^{i\alpha}\p_{A} W)
\nn &&\qquad\qquad\qquad\qquad+2\p_2(e^{-i\alpha}W-e^{i\alpha}\bar
W)\nn &=&-2{\rm Im}(e^{-i\alpha}W)\Big|^{x^2=\infty}_{x^2=-\infty}
\eea
where we have imposed the BPS equation.  We can rescale $Z^A \to \kappa^{-1/2} Z^A$ so that $\kappa$ only appears as an overall factor in the Lagrangian. Therefore the energy is proportional to $\kappa^{-1}$ and we can fix $\kappa$ by comparing the energy of a BPS fuzzy funnel with what is expected from the self-dual string (as was done in \cite{Basu:2004ed}).

Finally, we provide an example of a suitable non-associative algebra to make a more explicit connection with the  BPS equation of \cite{Basu:2004ed}.  We consider $N \times N$ matrices with a modified multiplication rule.  The trace form condition is quite restrictive; the simplest possibility that we found is
\be
A\cdot B  = QABQ,
\label{Qrule}
\ee
where $Q$ is some fixed, invertible matrix, and the usual matrix product is understood on the right-hand side. A suitable trace form is
\be
\Tr(A,B) = {\rm tr}(Q^{-1}AQ^{-1}B),
\ee
where tr denotes the usual matrix trace. The generalized Hermitian conjugate is $A^\# = QA^\dag (Q^{-1})^\dag$, where $\dag$ is the ordinary Hermitian conjugate. The associator in this algebra is
\be
<A,B,C>=Q^2ABQCQ-QAQBCQ^2.
\ee
To make contact with \cite{Basu:2004ed}, which contains a Hermitian matrix $G$ such that $G^2=1$, we take
\be
Q = \frac{1+iG}{\sqrt{2}}
\ee
so that $Q^2=iG$, $Q^\dag=Q^{-1}$ and $Q^\#=Q$.

The components of $Z^A$ that commute with $G$ associate with each other. Therefore we restrict to the subalgebra with $\{G,Z^A\}=0$. It follows that
\be
QZ^AQ = Z^A\qquad {\rm and} \qquad Z_A^\#=Z_A^\dag .
\ee
To construct the three-algebra, we take the vector space $\cal V$
spanned by $Q$ and all Hermitian matrices that anti-commute with
$G$.  The matrix $Q$ can be identified as the translation element
since
\be
[Q,A,B]=0,
\label{transis}
\ee
for all $A,B \in {\cal V}$.  It is not hard to show that
$(i[Z^A,Z^B,Z^C])^\# = i[Z^A,Z^B,Z^C]$ for all $A,B,C \in{\cal V}$.

With these definitions, the superspace action can be written as
\be {\cal L}  =  \frac{1}{2}\int d^4\theta\, {\rm tr}\,({\cal
Z}^A{\cal Z}^\dag_A) -\frac{\kappa}{8} \epsilon_{ABCD}\int
d^2\theta\,{\rm tr}\,(G{\cal Z}^A{\cal Z}^B{\cal Z}^C{\cal Z}^D) +
{\rm h.c.} \label{SSpace}\ee
The BPS equation is
\be {d Z_{\bar A}\over d(x^2)}+\frac{\kappa_1}{4!}
e^{i\alpha}\epsilon_{ABCD} \,[G,Z^B, Z^C,Z^D]=0.  \ee
This is nothing but the Basu-Harvey equation (\ref{BHeq}), provided that we take $\kappa_1 = M^3/8\pi\sqrt{2N}$ \cite{Basu:2004ed,Berman:2006eu}.  In fact it is slightly generalized since the $Z^A$ need not be real.
If we choose our solution so that $Z^B=Z_{\bar B}$, we also satisfy
the constraint $[Z^A,Z^B,Z_{\bar B}]=0$.

We note that the Lagrangian (\ref{SSpace}) has a global symmetry.  In particular $Z^A \to g Z^A g^{-1}$ (with the usual matrix product) is a symmetry provided $g \in U(N)$ and $[g,G]=0$.  However it is not clear to us that there is any significance to this symmetry because this algebra was chosen primarily for illustrative purposes.

\section{\sl Comments}

In this paper we studied the structure of the worldvolume
supersymmetry algebra of multiple M2-branes.  We
argued that it is natural for the embedding coordinates to take
values in a non-associative algebra.  We showed that the
supersymmetry transformations close into a novel gauge symmetry.  We
also presented a three-dimensional model Lagrangian for this system,
with reduced supersymmetry, containing only scalar and spinor
modes. The Basu-Harvey equation arises as the BPS condition for this
system.

A notable feature of $N$ M2-branes, as opposed to multiple
M5-branes, is that the number of degrees of freedom should scale as
$N^\frac{3}{2}$.  Since this is less than $N^2$, one might hope to
embed the algebra of the M2-brane coordinates into a matrix
representation.  The algebra constructed above is a twisted form of
the algebra of $N\times N$ matrices, based on some preferred matrix
$Q$, and as such seems to be fairly artificial. However, our setup
allows us to truncate our matrices to obtain a system with
$N^\frac{3}{2}$ degrees of freedom. Let us take $N = n^2$ and
consider the $N\times N$ matrices of the form
\be
X = \sum_{k=0}^{n-1} X_k\otimes\Omega^k ,
\ee
where $\Omega$ is an $n\times n$ matrix such that $\Omega^{n}=1$.
These matrices form a vector subspace of all $N\times N$ matrices
under the usual rule of addition and scalar multiplication. Thus $X$
consists of $n$, $n\times n$ matrices and hence has $n^3 =
N^\frac{3}{2}$ degrees of freedom. These matrices form an
associative algebra under matrix multiplication.  However, we can
also use the multiplication rule  (\ref{Qrule}), with $Q =
\Omega^q\otimes 1$ for some $q \in \mathbb{Z}$, to obtain a
non-associative algebra with $N^\frac{3}{2}$ degrees of freedom. In
particular if $n$ is divisible by $4$ then we reproduce the
Basu-Harvey equation by taking $q=n/4$ and identifying $G
=\Omega^\frac{n}{2}\otimes 1$.

This paper has been rather speculative in nature; there are many
outstanding issues that still need to be addressed. Perhaps the most
pressing is to further understand the local symmetries implied by
the full supersymmetry algebra. Also, our choice of non-associative
algebra was rather ad hoc; its main purpose was to illustrate how
things could work in principle. It would be of interest to find a
better motivated algebra to describe multiple M2-branes. In addition
it would be useful to make contact with the work of \cite{Bengtsson:2004nj}, which includes additional non-propagating fields on the brane worldvolume, and also to extend our results to obtain a manifestly U-duality invariant action. Last but not least, we hope and expect that the study of supersymmetric
theories and their solitons based on non-associative algebras will
prove to be fruitful in its own right.

\section*{\sl Acknowledgements}

We would like to thank D. Berman, N. Copland, C. Hofmnan and S. Ramgoolam
for helpful discussions. JB was supported by the National Science Foundation, grant NSF-PHY-0401513.  NL was supported in part by the PPARC
grant PPA/G/O/2000/00451 and the EU Marie Curie research training
work grant HPRN-CT-2000-00122.


\begin{thebibliography}{10}


\bibitem{Basu:2004ed}
  A.~Basu and J.~A.~Harvey,
  Nucl.\ Phys.\ B {\bf 713}, 136 (2005)
  [arXiv:hep-th/0412310].


\bibitem{Howe:1997ue}
  P.~S.~Howe, N.~D.~Lambert and P.~C.~West,
  Nucl.\ Phys.\ B {\bf 515}, 203 (1998)
  [arXiv:hep-th/9709014].



\bibitem{Constable:2001ag}
  N.~R.~Constable, R.~C.~Myers and O.~Tafjord,
  JHEP {\bf 0106}, 023 (2001)
  [arXiv:hep-th/0102080].


\bibitem{Diaconescu:1996rk}
  D.~E.~Diaconescu,
  Nucl.\ Phys.\ B {\bf 503}, 220 (1997)
  [arXiv:hep-th/9608163].

\bibitem{Kapustin:1998pb}
  A.~Kapustin and S.~Sethi,
  Adv.\ Theor.\ Math.\ Phys.\  {\bf 2}, 571 (1998)
  [arXiv:hep-th/9804027].


\bibitem{Tsimpis:1998zh}
  D.~Tsimpis,
  Phys.\ Lett.\ B {\bf 433}, 287 (1998)
  [arXiv:hep-th/9804081].




\bibitem{Guralnik:2000pb}
  Z.~Guralnik and S.~Ramgoolam,
  JHEP {\bf 0102}, 032 (2001)
  [arXiv:hep-th/0101001];
  S.~Ramgoolam,
  Nucl.\ Phys.\ B {\bf 610}, 461 (2001)
  [arXiv:hep-th/0105006];
  S.~Ramgoolam,
  JHEP {\bf 0210}, 064 (2002)
  [arXiv:hep-th/0207111].


\bibitem{Berman:2005re}
  D.~S.~Berman and N.~B.~Copland,
  Nucl.\ Phys.\ B {\bf 723}, 117 (2005)
  [arXiv:hep-th/0504044].


\bibitem{Nogradi:2005yk}
  D.~Nogradi,
  arXiv:hep-th/0511091.

\bibitem{Berman:2006eu}
  D.~S.~Berman and N.~B.~Copland,
  arXiv:hep-th/0605086.



\bibitem{Schwarz:2004yj}
  J.~H.~Schwarz,
  JHEP {\bf 0411}, 078 (2004)
  [arXiv:hep-th/0411077].

\bibitem{Lee:2006hw}
  S.~Lee,
  arXiv:hep-th/0610204.

\bibitem{Bengtsson:2004nj}
  V.~Bengtsson, M.~Cederwall, H.~Larsson and B.~E.~W.~Nilsson,
  JHEP {\bf 0502}, 020 (2005)
  [arXiv:hep-th/0406223].



\end{thebibliography}
\end{document}